# Interdependencies between Mining Costs, Mining Rewards and Blockchain Security[1]

*Pavel Ciaian*

*European Commission, Joint Research Centre (JRC)*

*Via E. Fermi 2749, 21027 Ispra, Italy*

*e-mail: pavel.ciaian@ec.europa.eu*

*d'Artis Kancs*

*(corresponding author)*

*European Commission, Joint Research Centre (JRC)*

*Via E. Fermi 2749, 21027 Ispra, Italy*

*e-mail: d'artis.kancs@ec.europa.eu*

*Miroslava Rajcaniova*

*Slovak University of Agriculture in Nitra (SUA) and University of West Bohemia (UWB)*

*Tr. Andreja Hlinku 2, 949 76 Nitra, Slovakia*

*e-mail: miroslava.rajcaniova@uniagr.sk*

[1] The authors gratefully acknowledge financial support received from the Slovak Research and Development Agency under the contract No. APVV-18-0512 and VEGA 1/0422/19. The authors would like to thank two anonymous referees for excellent suggestions as well as participants of the International Conference on Macroeconomic Analysis and International Finance as well as seminar participants at the European Commission for comments and useful suggestions. The conceptual framework of this paper is based on Ciaian et al. (2021). The authors are solely responsible for the content of the paper. The views expressed are purely those of the authors and may not in any circumstances be regarded as stating an official position of the European Commission.

**Interdependencies between Mining Costs, Mining Rewards and Blockchain Security**

**Abstract**

This paper studies to what extent the cost of operating a proof-of-work blockchain is intrinsically linked to the cost of preventing attacks, and to what extent the underlying digital ledger's security budgets are correlated with the cryptocurrency market outcomes. We theoretically derive an equilibrium relationship between the cryptocurrency price, mining rewards and mining costs, and blockchain security outcomes. Using daily crypto market data for 2014–2021 and employing the autoregressive distributed lag approach – that allows treating all the relevant moments of the blockchain series as potentially endogenous – we provide empirical evidence of cryptocurrency price and mining rewards indeed being intrinsically linked to blockchain security outcomes.

# 1 Introduction

Blockchain – a distributed network of anonymous record-keeping peers (miners) – is an inherently 'trustless' ledger. In the proof-of-work (PoW) blockchain, the trust problem among non-trusting parties is solved by requiring miners to pay a cost (in form of computing capacity) to record transaction information on a chain of blocks and requiring that other record-keepers (miners) validate those blocks. Mining incentives are ensured via rewards for a correct and secure record keeping – the reward for every block is allocated to the miner that first solves the computational problem (hash function) by using guess and check algorithms based on the new and previous blocks of transactions.

Cryptocurrency price shocks and hence changes in mining rewards (in a fiat currency denomination) affect mining incentives for the ledger record-keepers and hence the underlying ledger's security. Blockchain users – who value the network security – in turn adjust their crypto coin portfolio exercising in such a way upward or downward pressure on cryptocurrency's price (see Figures 2 and 3). The literature suggests that these interdependencies between cryptocurrency's value, mining costs and blockchain security may contribute to the extreme volatility of cryptocurrency return (Ciaian et al. 2021b; Pagnotta 2021).

The present paper studies these interdependencies between mining costs, mining rewards and blockchain immutability. We attempt to answer the following questions. To what extent the cost of operating blockchains is intrinsically linked to the cost of preventing attacks? How closely interrelated are the digital ledger's record-keeping security budgets (measured by mining rewards in a fiat currency nomination) of cryptocurrencies with the cryptocurrency market outcomes? We focus on the proof-of-work blockchain, which is a particularly interesting blockchain to study as the involved physical resource expenditures provide a distinct advantage in achieving consensus among distributed miners. Our results suggest that the cryptocurrency price and mining rewards are indeed intrinsically linked to blockchain security outcomes, the elasticity of mining rewards being higher than that of mining costs with respect to the network security equilibrium.

The previous literature has mostly studied blockchain security concerns from a crypto-coin user perspective (see Lee 2019, for a survey). It has found that crypto-coin users value the underlying ledger security, they internalize and price the risk of a blockchain attack that could compromise the ability to exchange crypto-coins for goods. Blockchain users who engage in on-chain transactions – only the on-chain transactions are secured by the mining rewards – value security measured by the amount of computational power committed to the blockchain; ceteris paribus they prefer more computing power being committed to the ledger. There is little empirical evidence, however, available in the literature about the interdependencies between mining costs, mining rewards and blockchain security. Moreover, there is confusion in the literature that the blockchain security would be an embedded property of the underlying institutional governance's technology.

Our main contribution to the literature is formally establishing a link between the probability distribution over security outcomes depending permanently on the underlying distribution of cryptocurrency market outcomes and providing a supporting empirical evidence. The papers most closely related to ours are from the theoretical literature on the blockchain mining and security (Iyidogan 2020; Pagnotta 2021; Ciaian et al. 2021a). In particular, our paper is related to the emerging literature on the economic properties and implications of blockchains (Abadi and Brunnermeier 2018; Budish 2018; Biais et al. 2019). This literature studies coordination among miners in a blockchain-based system and shows that while the strategy of mining the longest chain is in fact an equilibrium, there are other equilibria in which the blockchain forks, as observed empirically. Whereas Abadi and Brunnermeier (2018) place most of the focus on coordination among users; record-keepers' payoffs are determined by users' actions, and a global games refinement of the game played among users puts more discipline on exactly how and when a fork may occur, in Biais et al. (2019) forks occur for several reasons and are interpreted as causing instability; record-keepers' payoffs when forking depend exogenously on the number of record-keepers who choose a given branch of the fork. Budish (2018) studies the costs of incentivizing honesty for cryptocurrency blockchains. Cong and He (2019) focus mostly on the issue of how ledger transparency leads to a greater scope for collusion between users of the system. An alternative perspective studied in the literature is the collusion between the blockchain's record-keepers rather than between users, which shows that collusion can occur only when entry of record-keepers is constrained.

Our results complement the findings of this emergent literature by quantifying how the probability distribution over security outcomes permanently depends on the underlying distribution of cryptocurrency market outcomes. Due to the extremely high cryptocurrency volatility, also the PoW-blockchain security budget is exposed to high volatility and may result in a series of low-security equilibriums and high-security equilibriums. In contrast, the physical resource cost to write on the blockchain – the cost of operating the PoW-blockchain – is only weakly cointegrated with the strength of the network security. This cointegration relationship is geographically differenced – it is more significant for the world global mining leader China than for other world regions. The mining cost effect seems to trigger a downward pressure on the extensive margin of mining more in North America and Europe than in China, where the increasing intensive margin of mining more than offsets the negative effects on the network hash rate and hence the blockchain security. The ARDL estimates for the speed of adjustment of the PoW-blockchain security suggest that after temporary shocks to crypto markets any disequilibria is corrected and the security equilibrium reverts back to mean in the long-run.

The paper proceeds as follows. First, we provide a background literature overview about the PoW-blockchain (section 2). Second, we establish a theoretical relationship between the cost of proof-of-work, cryptocurrency market outcomes and blockchain security outcomes (section 3). Third, we estimate empirically the derived structural PoW-blockchain security relationships for cryptocurrencies that rely on application-specific integrated circuits (ASICs), such as Bitcoin. We use daily Bitcoin data for 2014–2021 and employ an autoregressive distributed lag approach that allows treating all the relevant moments of the blockchain series as potentially endogenous (sections 4 and 5). To examine the extent to which this relationship is contingent upon exogenous price shocks, the role of the cryptocurrency mining reward and the proof-of-work cost for each of the respective moments is estimated after accounting for the information embedded in the lags of the entire distribution of blockchain security outcomes. The main results are presented in section 6. They suggest that the cryptocurrency price, mining rewards and mining costs are intrinsically linked to blockchain security outcomes. The final section concludes.

## 2 The record-keeping of digital transactions

### 2.1 Blockchain and the proof-of-work

Blockchain is a distributed alternative to centralized transaction-recording and record-keeping systems by enabling trustworthy interactions, recording transactions among non-trusting parties and storing interaction records. The underlying ledger that creates and stores records of transactions is a digital chain of blocks, where information is recorded sequentially in data structures known as 'blocks' stored into a public database ('chain'). Being distributed, blockchain is run by a peer-to-peer network of nodes (computers) who collectively adhere to an agreed distributed validation algorithm (protocol) to ensure the validity of transactions. A distributed network of anonymous record-keeping peers (miners) with free entry and exit is inherently 'trustless' and thus requires a trust-enhancing mechanism. To solve the trust problem among non-trusting parties, miners are obliged to pay a cost (in form of computing power for blockchain) to record transaction information and requiring that future record-keepers (miners) validate those reports. Under a well-functioning institutional governance technology, blockchain is immutable, meaning that once data have been recorded on the blockchain, it cannot be altered.

The key preconditions for a well-functioning market are accuracy and security of transactions and enforcing property rights and contracts. In traditional centralized institutional governance systems, usually, state or other type of centralized authority (intermediary) guarantees the transfers of ownership, ensures transfers of possessions, guarantees the security of property rights and contract enforcement. The honest behavior of the centralized intermediary is incentivized through monopoly rents. A comparative advantage of distributed institutional governance systems such as blockchain is the ability to achieve and enforce a uniform view (agreement) among non-trusting parties with divergent interests and incentives on the state of transactions in a cost-efficient and consensus-effective way. Blockchain security algorithms make it possible for distributed record-keepers to ensure that the network rules are being followed, i.e. all other record-keepers disregard any chain containing a block that does not conform to the network rules. The correctness and security is incentivized via physical resource costs: proof-of-work (PoW) makes it costly to extend invalid chains of blocks (Davidson, De Filippi and Potts 2016; Cong and He 2019).[2]

Given its cost-efficiency and consensus-effectiveness advantages, blockchain's potential applications go far beyond the creation of decentralized digital currencies, it can be used to achieve consensus of virtually any type of records or transactions, particularly of those related to property rights, transfer of property rights and contract execution (Davidson, De Filippi and Potts 2016). Blockchain technology has the potential to serve as an irreversible and tamper-proof public record repository for documents, contracts, properties, and assets as well as it can be used to embed and digitally store information and instructions with a wide range of applications. For instance, smart contracts (self-executing actions in the agreements between two or multiple parties), smart properties (digitally recorded ownership of tangible and intangible assets) or decentralized autonomous organizations (DAOs) (Atzori 2017). Such a system might sustain various activities spanning from financial transactions, identity management, data sharing, medical recordkeeping, land registry up to supply chain management and smart contract execution and enforcement. Blockchains can also record obligations; distributed ledgers could be used in the fintech space to track consumers' transactions and credit histories (e.g. Davidson, De Filippi and Potts 2016; Nascimento, Pólvora and Sousa-Lourenço 2018).

In the same time, ensuring a transaction correctness and security may be more challenging for distributed digital ledgers than for traditional centralized ledgers (Abadi and Brunnermeier 2018).[3] First, because digital goods are non-rival and non-excludable, which unlike traditional private goods do not prevent a double spending. Second, the security budget of distributed ledgers is endogenous and fluctuates over time (in a fiat currency nomination – see Figure 1), implying that the underlying institutional governance technology may become vulnerable to attacks during cryptocurrency's low-price low-security-budget periods. Hence, marinating the correctness and security of transactions may become a challenge especially in periods of low security budget. Indeed, a number of cryptocurrency-blockchains with a comparably small security budget of preventing attacks have experienced successful majority (hash rate)[4] attacks in recent years, e.g. Bitcoin Gold, Ethereum Classic.[5]

---

[2] There are two main types of validation mechanism – proof-of-work (PoW) and proof-of stake (PoS) – with each having different incentive scheme in achieving consensus. This paper focuses on the PoW linked to Bitcoin which is the largest and most popular cryptocurrency.
[3] A strong security of information in the context of a distributed ledgers implies immutable records of transactions, including ownership rights and smart contracts.

**2.2 Blockchain mining**

The blockchain mining consists of nodes (called miners or record-keepers) of a distributed network competing for the right to record sequentially information about new transactions to the digital ledger. In the case of PoW, miners have to solve a computationally challenging problem in order to record information and validate others' records on the ledger (in intervals of around ten minutes in Bitcoin). Solving the computational problem (puzzle) is energy intensive and thus costly. First, miners have to invest in a computing capacity; these costs are fixed and independent of the success rate. Second, miners have to incur variable costs, such as energy (and time) for the computationally-intense mining process, and rental expenses for the location of the mining equipment. On the revenue side, mining incentives are ensured via rewards for a correct and secure record keeping. The reward for every block is allocated to the miner that first solves the computational problem (hash function), by using guess and check algorithms based on the new and previous blocks of transactions. The winning miner broadcasts both the new block of transactions and the solution to the computational problem to the entire decentralized network; all other network participants "*express their acceptance of the* [new] *block by working on creating the next block in the chain, using the hash of the accepted block as the previous hash*" (Nakamoto 2008).

The miner's computing capacity is the main mining input, its performance is measured in a hash rate, which measures the speed at which a given mining machine operates. Usually, the hash rate is expressed in hashes per second (h/s). For example, a mining machine operating at a speed of 100 hashes per second makes 100 guesses per second. Thus, the hash rate measures how much computing capacity blockchain is deploying to continuously solve the computational problem and generate/record blocks. Given that a mining computer has to make many guesses to solve the computational problem; higher hash rate allows a miner to have higher number of guesses per second, thus increasing his/her chance to first solve the computational problem and receive the reward.

The computational problem to be solved by miners adjusts endogenously, depending on the number of network participants and the aggregate blockchain computing capacity it is adjusted to become more difficult or less difficult. Bitcoin's mining (hashing) difficulty algorithm is designed to adjust after every 2 016 blocks (approximately every 14 days) to maintain an interval of approximately 10-minutes between blocks. When the aggregate blockchain computing capacity increases, the computation problem difficulty adjusts upwards (i.e. the required hash rate to 'mine' a block increases), whereas in periods of a low mining network participation, it decreases. The adjustment in the mining difficulty level is done for the purpose to compensate / counterbalance changes in the aggregate blockchain computing capacity employed by miners (Joudrey 2019; BitcoinWiki 2021).

---

[4] The hash rate measures the speed at which a given mining machine operates. The hash rate is expressed in hashes per second (h/s) (or number of guesses per second), which measures how much computer capacity a cryptocurrency network is devoted to solve the computational problem and generate/record blocks.
[5] For example, Bitcoin Gold, a hard fork of Bitcoin, was subject to several double-spending attacks in 2018 causing a price reduction (in USD) by around 40%. Similarly, Ethereum Classic also experienced a double-spend attack in 2019 and 2020 also leading to its price decrease.

The network hash rate also determines the security and stability of the underlying blockchain institutional governance technology (Figures 2 and 3). The physical resource cost to write on the PoW-blockchain – the cost of operating the blockchain – is intrinsically linked to the cost of preventing attacks.[6] Higher hash rate implies stronger security, because any dishonest miner (attacker) would need to employ more resources (computing capacity) to attack the institutional governance technology of blockchain.[7] In the context of creating and maintaining distributed ledgers of information, a strong security implies immutable records of transactions, including ownership rights and smart contracts.

**2.3 Digital goods and decentralized ledgers vis-à-vis physical goods and centralized legers**

*Correctness and security*. The system security reflects the probability of an attack; security is paramount to any financial or non-financial network since transfers of ownership and enforcement of property rights require verifications, and it should be difficult for an attacker to manipulate historical or/and new records. In a centralized system, a specific trusted agent assumes such responsibility. In blockchain, however, verification and updates to the system ledger rely on self-selected non-cooperating agents – miners. In cryptocurrencies such as Bitcoin, the reward to successful miners includes transaction fees and a predetermined number of newly minted bitcoins, which role is to incentivize miners to devote computing capacity for block validation and thus to provide the system's security. The probability of an attack is driven by the balance of a computing power between potential attackers and honest miners. Greater number of (honest) miners and more computing capacity imply smaller probability of a successful attack (Figure 2).

---

[6] For illustration purpose, an example of a majority attack and gain from double-spending as provided by Van Valkenburgh (2018) could look as follows: An attacking miner with the majority computing power compiles a secret (private) version of the Bitcoin blockchain. At the same time the attacking miner sends, for example, 100 Bitcoins to a Bitcoin exchange, sells them and sends the received money (dollars) to his/her bank account. This Bitcoin transaction is incorporated into the public blockchain run by honest miners. The exchange observes the transaction on the public (honest) blockchain, thus assumes it has the 100 Bitcoins and initiates money transfer to the attacker's bank account. However, the attacker does not send the 100 Bitcoins to the exchange in his/her own secret blockchain version. Once the attacker receives the money to his/her bank account, the private version of the Bitcoin blockchain can be broadcasted to the network. Because the attacker has more computing power than the rest of the network combined, the attacker private chain will be longer (more cryptographic problems solved) and the rest of the network will recognize this new blockchain as the valid one. According to the new reorganized chain, the exchange that accepted the 100 Bitcoins for money no longer has those 100 Bitcoins as well as it lost their dollar value of Bitcoins which were sent to the attacker bank account. In contrast, the attacker has both the 100 Bitcoins and the dollar value of 100 Bitcoins (Van Valkenburgh 2018).

[7] Different types of blockchain attacks include selfish mining, the 51% attack, Domain Name System (DNS) attacks, distributed denial-of-service (DDoS) attacks, consensus delay (due to selfish behavior or distributed denial-of-service attacks), blockchain forks, orphaned and stale blocks, block ingestion, smart contract attacks, and privacy attacks.

Compared to traditional centralized ledgers, ensuring a transaction correctness and security may be more challenging for distributed digital ledgers like Bitcoin, because digital goods are non-rivalry and non-excludability which compared to traditional private goods do not prevent a double spending, and the security budget of distributed ledgers is endogenous and fluctuates over time (i.e. if the value mining rewards changes in a fiat currency nomination). Miners react to expected profit incentives by adjusting their computing capacity. For example, low expectations of a cryptocurrencies price would induce reducing computing capacity devoted to mining, thus rendering the network more vulnerable and potentially further magnifying the cryptocurrency's price decrease. Due to miners' rational responses, the realization of pessimistic cryptocurrency's market outcomes also implies that a cryptocurrency's security can severely worsen resulting in a low-security equilibrium and lowering the network's life expectancy as measured by the average time until a successful attack (Pagnotta 2021).

Double-spending attacks are one of the largest security concerns among blockchain users. Cryptocurrencies that have a relatively small security budget of preventing attacks have experienced a number of successful majority hash rate attacks in recent years. For example, Bitcoin Gold, a hard fork of Bitcoin, experienced a sequence of double-spending attacks in May 2018. Its price measured in USD at the end of that month was 40% lower. Ethereum Classic also experienced a double-spend attack and several deep block reorganizations, following a 50% decline in its price and hash rate in January 2019. Double-spending attack is also possible when the blockchain in question handles assets other than currency. For example, a financial institution that loses money on a trade may wish to reverse the history of transactions including that trade.

*Competition and costs.* Blockchain differs from centralized legers (e.g. notary offices, cadastral offices, banks) along several dimensions of the market structure, competition being one of them. Typically, centralized ledgers are managed by monopolists (e.g. central banks) that extract distortionary rents from the ledger's users, because the entry is not free and switching between legers is costly for users. Traditional centralized record-keeping systems provide incentives to record honestly by monopoly profits and the fear to lose the future monopoly rents (Abadi and Brunnermeier 2018). In contrast, distributed record-keeping system allow for competition: there is a free entry (every miner can write on the ledger, subject to network rules) and switching between ledgers (e.g. 'forks') is costless for users.[8] The blockchain market structure with free entry and fork competition eliminates the rents that a monopolist would extract in an identical market and eliminates the inefficiencies arising from switching costs in centralized record-keeping systems (Huberman, Leshno, and Moallemi 2019). There is also competition between a potential attacker and all honest miners. Incentives to record honestly – that are provided through the imposition of a physical resource cost to write on the blockchain – make it costly for a potential attacker to distort the ledger.

---

[8] For example, a hard fork preserves all of the data in the parent blockchain: e.g. Bitcoin Gold and Bitcoin Cash in the case of hard forks of Bitcoin.

Free entry and competition ensure that distributed ledgers can be more efficient and transactions less costly than centralized ledgers. Blockchain miners can enter freely, meaning that any agent who wishes to write on the ledger may do so by following an agreed set of rules. However, free entry of anonymous record-keepers is 'trustless' and thus requires a trust-enhancing mechanism. Public blockchains typically solve the trust problem by forcing record-keepers to pay a physical resource cost to record information and requiring that future record-keepers validate those reports. In the case of the proof-of-work (PoW), miners have to solve a computationally challenging problem in order to record information and validate others' reports. The physical resource cost to write on the blockchain is the main the cost of operating a blockchain. Compared to distortionary rents of centralized ledgers, in distributed blockchain-based record-keeping systems, welfare losses stem mainly from the waste of computational resources, computing and electricity costs of mining. Enforcing the execution of transactions by a cryptographic code ensures a significant reduction of transaction costs. According to Huberman, Leshno, and Moallemi (2019), the physical resource costs of competing non-cooperating miners are significantly lower than monopoly rents of centralized ledgers. Hence, an important cost advantage of the blockchain technology compared to centralized record-keeping systems consists of avoiding centralized intermediaries (e.g. a notary, cadastral office, banks) and the associated rents.

*Network externalities.* Digital distributed ledgers such as blockchain are subject to network externalities, which are not present under centralized ledgers. When miners engage in the mining of blockchains, both positive and negative network externalities related to the blockchain security emerge. The positive network externality suggests higher blockchain security as the number of miners increases, because each additional node strengthens the chain's security, by making it more difficult for any individual miner to launch an attack or to guess who will be the winning miner (Waelbroeck 2018). The negative network externality occurs because each individual miner invests in the mining-computing power, which increases both the individual miner's marginal income though also mining costs, as the difficulty of the computational problem increases in the number of miners and their computing power ("hash-power"). Subsequently, higher difficulty of mining reduces the incentives for mining and increases the concentration of mining activities, as miners are learning by mining, resulting in reduced blockchain security (Parra-Moyano, Reich and Schmedders 2019). When many small miners enter the blockchain network, likely, the positive network externality will dominate and the blockchain security outcomes will be superior compared to a highly skewed distribution of computing power across miners (few mining pools having a large share of the total network hash rate).

Individual non-cooperating agents do not internalize these network externalities when making their optimal decisions. Blockchain users (agents who engage in transactions) take the price and security levels as given and, unlike the central planner, do not internalize the impact of their decisions on mining costs. Similarly, miners do not internalize the effect of their hash rate choice on the blockchain security. For equilibria that display high security levels, given the decreasing security gains from a mining investment, the part that miners fail to internalize is decreasing. In contrast, the part that blockchain users fail to internalize is not decreasing, because marginal mining costs are not decreasing (Pagnotta 2021).

Fourth, being a distributed *institutional governance technology* for creating and maintaining distributed ledgers of information, it is different from a centralized institutional governance. According to Davidson, De Filippi and Potts (2016), blockchain provides a new "institutional technology" or a "governance technology". "[Blockchain can be understood] *as a revolution (or evolution) in institutions, organization and governance*" (Davidson, De Filippi and Potts 2016). Compared to traditional centralized intermediaries, digital rules are enforced by a distributed network of interconnected non-trusting parties. Trustworthily interactions between non-trusting parties are executed and recorded on a distributed network by eliminating the need for a centralized intermediary. A consensus mechanism ensures that the true history is recorded on the ledger, rejecting fraudulent records. The build-in validation processes in the PoW consensus algorithm and the use of cryptographic signatures and hashes ensures the network governance, disincentivizes dishonest nodes to insert fake or malformed transactions in the blockchain, and ensures the trustworthiness of transactions among non-trusting parties on blockchain. The computationally established trustworthiness of the institutional governance technology ensures accuracy in establishing, delineating and protecting ownership rights (i.e. it allows owners to exercise ownership rights in terms of use, transfer, or exploitation of assets); it can execute and enforce contracts (through smart contracts); and it can encompass various types of organizations through DAOs (e.g. firms, venture capital funding, non-profit organization).

Following North (1990), institutions are "*the rules of the game in a society*" and includes both formal rules such as laws and informal constraints such as "*codes of conduct, norms of behavior, and conventions*". Formal rules are enforced by state, while informal rules are enforced by the members of the relevant group (North 1990; Kingston and Caballero 2009; Greif and Kingston 2011). According to Hodgson (2006) "*institutions are systems of established and embedded social rules that structure social interactions*". From this point of view, blockchain is a type of distributed (informal) institution with digitally embedded rules of the game – defined within the validation algorithm and enforced through a decentralized network of participants – that structure digitally recorded interactions between agents. According to Davidson, De Filippi and Potts (2016) "[blockchain is] *an 'institutional technology', a governance technology for making catallaxies, or rule-governed economic orders. Blockchains thus compete with firms, markets and economies, as institutional alternatives for coordinating the economic actions of groups of people, and may be more or less efficient depending upon a range of conditions (behavioural, cultural, technological, environmental, etc).*"

Fifth, significant changes in the *global technological development* (e.g. new/faster technologies become available) or macroeconomic environment, require adjustments in the institutional governance – either by a central authority or endogenously. Technological changes underlying digital institutional governance systems are considerably faster compared to traditional institutional governance systems. In traditional centralized record-keeping systems, the frequency of important technological changes and the required institutional governance adjustments is low; the key role in adjusting institutions to changes in the external environment plays the centralized intermediary. In digital self-enforcing record-keeping systems, the frequency of technological changes and the required institutional governance adjustments is high; the institutional governance is adjusted endogenously. For example, the blockchain network governance employs the PoW consensus algorithm and uses cryptographic signatures and hashes. The institutional governance technology of Blockchain is frequently adjusted (every 14 days) to changes in the technological development (e.g. growth of the mining processor computing speed) or macroeconomic environment (e.g. significant increase in the cryptocurrency's value and hence mining rewards in a fiat currency denomination) by adjusting mining incentives for the network record-keepers (Dollar and Kraay 2003; Hodgson 2006; Glaeser et al. 2004; Kingston and Caballero 2009; Greif and Kingston 2011).

Blockchain provides a particularly interesting case to study, as the institutional governance system (i.e. the security of the enforcement of property rights and contracts) is determined endogenously by the underlying governance technology. Whereas traditional centralized institutional governance systems often are path-dependent and face many impediments to bring about evolutionary developments (e.g. interest group pressure, bargaining and political conflict between interest groups) which makes them less sensitive to economic changes (North 1990; Kingston and Caballero 2009; Greif and Kingston 2011), the flexibility in institutional adjustment of digital distributed ledgers allow them to adopt and accommodate changing market conditions such as integrating and stimulating the growth of new technologies and "non-tangible" innovations. Institutional rigidity (neutrality) could be desirable in certain situations particularly when underlying factors (e.g. economic crisis, political crisis, civil conflicts, wars) pressure towards lower-quality institutions, but not in others (uni-directional institutional change towards its improvement is desirable, not vice versa). Hence, it is important to understand the interdependencies between mining costs, mining rewards and blockchain security.

## 3 Conceptual Framework

### 3.1 The model

We want to determine theoretically the equilibrium relationships between blockchain security, cryptocurrency market outcomes and resources devoted to the blockchain mining. Building on the mining models of Thum (2018) and Budish (2018) and considering the PoW of the most popular cryptocurrencies as example, we model a rational miner *i* that decides on the quantity of computing capacity, $m_{it}$ (e.g. expressed by the number of computer operations), to devote for mining each block *t* (represented in block time measured in 10 minute interval which is the average time needed to mine a block in blockchain). The mining output is measured in capacity of blockchain security units.

The probability of miner *i* winning the contest (i.e. the right to generate a new block and collect reward) depends on his/her computing capacity devoted for each block relative to the computing capacity of other miners. Previous studies assume that the probability of winning the contest and validating a block is independent of the miner size: $m_{it}/(m_{it}+\sum_{j\neq i}^{n_t} m_{jt})$, where $n_t$ is the total number of miners and $\sum_{j\neq i}^{n_t} m_{jt}$ is the total blockchain computing capacity of other miners devoted to the block *t* (e.g. Cocco and Marchesi 2016; Thum 2018). However, Parra-Moyano, Reich and Schmedders (2019) show that the probability of relatively bigger miners winning the mining contest is higher than that of relatively smaller miners because there is a "learning" effect when mining a particular block with larger mining computers learning faster than smaller mining computers. To account for the *learning by mining*, we assume the following transformation of the probability for a miner winning a block: $e^{m_{it}^{\gamma}}/(e^{m_{it}^{\gamma}}+\sum_{j\neq i}^{n_t} e^{m_{jt}^{\gamma}})$, where $\gamma$ is a transformation parameter (with $0 < \gamma \leq 0$), which implies that the ratio of odds between big and small miners (mining computers) of winning a block increases with the miners' size, $m_{it}$, while keeping the ratio of miner' size between miners fixed.

The purchase price of one unit of a computer equipment of a given efficiency, $\varepsilon$, is denoted by $q_t$. The successful miner receives reward $p_t R_t$, where $R_t$ is cryptocurrency quantity and $p_t$ is the cryptocurrency price per one unit expressed in monetary values (e.g. US dollar). Miner *i* chooses computing capacity, $m_{it}$, for a given computer efficiency, $\varepsilon$, so that to maximize the present discounted value of the flow of profits over the infinite time horizon:

(1) $$\pi_i = \sum_t \left(\frac{1}{1+\rho}\right)^t \left(\frac{e^{m_{it}^{\gamma}}}{e^{m_{it}^{\gamma}}+\sum_{j\neq i}^{n_t} e^{m_{jt}^{\gamma}}} E(p_t) R_t - c_t m_{it} - q_t I_{it}\right) - F$$

Subject to $m_{it+1}$ units of computing capacity:

(2) $$m_{it+1} = (1-\delta)m_{it} + I_{it}$$

where $c_t$ denotes variable costs per computer operation (e.g. energy cost), $E(p_t)$ is the expected cryptocurrency price, $\delta$ is depreciation rate, $I_{it}$ is investment in computer equipment, $F$ are one-time fixed costs (e.g. building – see Garratt and van Oordt 2020), and $\rho$ is a discount rate for time preference. Deviations from the expected price are random shocks, $v$, with an expected value of zero: $E(p_t) = p_t^*$, where $p_t = p_t^* + v$. We assume a rational price expectation framework of Muth (1961) in which miners base their cryptocurrency price formation on all the available information at the time when making their decisions on the investment in $m_i$. Miners are identical, risk-neutral, non-cooperative and profit-driven agents that invest according to the anticipated real value of block rewards.

Maximizing miner $i$'s profits for the given blockchain computing capacity of all other miners yields the following optimal conditions:

(3) $$-q_t + \frac{1}{1+\rho}\lambda_{t+1} = 0$$

(4) $$\frac{\gamma m_{it}^{\gamma-1} e^{m_{it}^\gamma} \sum_{j \neq i}^{n_t} e^{m_{jt}^\gamma}}{\left(e^{m_{it}^\gamma} + \sum_{j \neq i}^{n_t} e^{m_{jt}^\gamma}\right)^2} E(p_t)R_t - c_t + \frac{1}{1+\rho}(1-\delta)\lambda_{t+1} = \lambda_t$$

(5) $$m_{it+1} = (1-\delta)m_{it} + I_{it}$$

where $\lambda_t$ is a shadow price for a unit of computer resources.

Assuming a steady state equilibrium with $m_{it} = m_{il}$, $R_t = R_l$, $E(p_t) = E(p_l)$, $q_t = q_l$, and $n_t = n_l$ for $t \neq l$ and a symmetric equilibrium with $m_{it} = m_{jl}$, the equilibrium computing capacity per miner can be derived from equations (3) to (5) as follows:

(6) $$m_t = \left[\frac{\gamma(n_t-1)}{n_t^2} \frac{E(p_t)R_t}{c_t + (\rho+\delta)q_t}\right]^{\frac{1}{1-\gamma}}$$

Rewriting equation (6) in terms of the total blockchain computing capacity devoted to mining, $n_t m_t^*$, yields the mining equilibrium:

(7) $$n_t m_t = \left(\frac{1}{n_t}\right)^{\frac{1+\gamma}{1-\gamma}} \left[\gamma(n_t-1)\frac{E(p_t)R_t}{c_t + (\rho+\delta)q_t}\right]^{\frac{1}{1-\gamma}}$$

Equation (7) implies that the total blockchain computing capacity increases in the relative gain from mining, $E(p_t)R_t/(c_t + (\rho+\delta)q)$. The mining equilibrium implies that the blockchain computing capacity devoted to mining fluctuates with the cryptocurrency price. This model feature reflects the intuition that, ceteris paribus, higher nominal reward or higher cryptocurrency price induces miners to invest in more computing capacity. The opposite is true when agents anticipate the value of cryptocurrency to be low, miners have little incentive to invest in computational resources, and the security of the network is low.

Second, the mining equilibrium (7) implies that the total blockchain computing capacity increases at a decreasing rate in the level (intensity) of competition, $(n_t-1)/n_t^2$.

Third, equation (7) implies that miners have incentives to revert to the equilibrium level of the blockchain computing capacity as a response to cryptocurrency price shocks because otherwise miners would experience losses.

We follow Abadi and Brunnermeier (2018) and assume a free entry equilibrium where miners enter until profits are driven to zero. In the blockchain system, miners don't compete in prices but in capacity, similar to Cournot-type firms. An increase in the processing power of competing miners results in the expansion of the total blockchain computing capacity. In the presence of network externalities, free entry of miners serves to pin down the strength of the security.

Using equations (5) and (6), it is possible to derive the equilibrium number of miners, $n_t$, depending on mining returns, variable costs, fixed costs and the level (intensity) of competition, $(n_t - 1)/n_t^2$:

$$(8) \qquad n_t = E(p_t)R_t / \left( \rho F + (c_t + \delta q_t) \left[ \frac{\gamma(n_t-1)}{n_t^2} \frac{E(p)R}{c_t+(\rho+\delta)q_t} \right]^{\frac{1}{1-\gamma}} \right)$$

Fixed costs are related to credit constraint and rigidities to increase capacity related to financing the entry costs into the mining.

Equations (3) to (7) define the equilibrium behavior of honest miners by pinning down the level of computer resources they would allocate for mining at a given level of reward and competition from other miners. The total blockchain computing capacity devoted to the blockchain mining, $n_t m_t$, determines the security of blockchain. As discussed above, the more challenging is the computational mining puzzle to solve, the safer and more stable is the institutional governance technology because it becomes more costly for a potentially dishonest miner to conduct an attack. Such an attack may adversely affect the perception of cryptocurrency by its users reducing their trust and hence valuation of the cryptocurrency. If the reduction of the trust is large, it may cause a collapse in the economic value (price) of cryptocurrency. As equation (7) implies, the blockchain computing capacity for mining and hence the hash rate of the network would reduce, which might eventually lead to a collapse of blockchain. Thus, the security of the PoW blockchain depends on the size of mining reward received by miners which also determines the total blockchain computing capacity determined in equation (7).

### 3.2 Blockchain security and attacks

The probability of a (successful) attack on blockchain is reflected in the underlying ledger's security, it is inversely related the blockchain's security budget. This probability is driven by the balance of computing power between an attacker and honest miners. As noted by BitcoinWiki (2021), "*Bitcoin's security model relies on no single coalition of miners* *controlling more than half the mining intensity*".[9] A miner who controls more than 50% of the total blockchain computing capacity could exercise attack on blockchain that involves the addition of blocks that are somehow invalid or reverse previous accepted transactions ("majority attack"). Either the blocks contain outright fraudulent transactions, or they are added somewhere other than the end of the longest valid chain. A successful majority attacker could prevent (for the time that the attacker controls mining) confirmation of new transactions (e.g. by producing empty blocks) and reverse own transactions which potentially allows double-spending thus affecting all transactions that share the history with reversed transactions (BitcoinWiki 2021).

In our model, to control a majority power, equation (7) implies that an attacker must control more than 50% of the total blockchain computing capacity, $An_t m_t$, where $A > 1$. If we assume that the attack takes the duration equal to *s* block time, then the attacker's costs[10] are $sA(cn_t m_t + q_t n_t I_t) - (1 - \theta)qn_t m_t$ and the mining reward during the attack is $sp_t R_t$, where $\theta$ ($0 \leq \theta \leq 1$) represents the proportion of the mining technology, $m_t$, that can be recovered (reused, resoled, repurposed) after the attack.[11] The first term of the attacker's costs, $sA(c_t n_t m_t + qn_t I_t)$, includes energy and investment costs, while the second term, $(1 - \theta)qn_t m_t$, represents the loss related to the part of mining technology that cannot be recovered after the attack.

To des-incentivize and deter attacks on blockchain, the cost of an attack must be greater than the potential gain from an attack. Using the optimal condition (5), this implies the following incentive compatibility condition for blockchain against attacks:

(9) $$sAnm^*[(c_t + q_t\delta) - (1 - \theta)q_t] \geq (1 - \Delta)sE(p_t)R_t + V_A(\Delta)$$

where $\Delta$ ($0 \leq \Delta \leq 1$) is the proportional decrease in the price of cryptocurrency after the attack and $V_A$ is the expected payoff of the attack which is dependent on $\Delta$ and is equal to the sum of gains, $V_t(\Delta)$, obtained over the duration of attack *s* with $V_A(\Delta) = \sum_s V_t(\Delta)$.[12] The payoff from the attack, $V_A$, can represent the gain from a cryptocurrency double-spending or other type of gains (e.g. gain from a short sale of cryptocurrency, gain in cryptocurrency future markets from price fluctuation caused by the attack).

Using equation (6), the incentive compatibility condition (9) can be rewritten as:

---

[9] Although Bitcoin has not suffered from a majority attack, a number of Altcoins were subject to successful attacks in the past. For example, this was the case of the Bitcoin hard fork (Bitcoin Gold) in May 2018 (stealing $18 million worth of Bitcoin and other cryptos), Ethereum Classic (ETC) in January 2019 (double spending to over 200,000 ETC worth around $1.1 million), and Verge (XVG) was attacked several times in 2018 (with the biggest attack extracting about 35 million of XVG) (ViewNodes 2019).
[10] According to Crypto51 (2021), the theoretical cost of a 51% attack on Bitcoin is $ 413,908 per one hour.
[11] Note that if Bitcoin does not collapse after the attack, the mining equipment can be reused in continuing mining Bitcoin.
[12] Note that in the steady state situation assumed in the incentive compatibility condition (9), implies that the discount rate $\rho$ cancels out with $V_t(\Delta) = V_l(\Delta)$ for $t \neq l$.

(10) $$\left\{[E(p_t)R_t]^{\frac{\gamma}{1-\gamma}} - \frac{(1-\Delta)}{\beta}\right\} sE(p_t)R_t \geq \frac{V_A(\Delta)}{\beta}$$

where $\beta = An_t[(c_t+q\delta)-(1-\theta)q] \left[\frac{\gamma(n_t-1)}{n_t^2} \frac{1}{c_t+(\rho+\delta)q_t}\right]^{\frac{1}{1-\gamma}}$

Consider an attack where the only gain, $V_A$, is double spending. The attacker acquires $X$ units of crypto coins which (s)he double spends during the attack by exchanging them for the standard fiat currency. This implies that the gain from attack is $V_A(\Delta) = E(p_t)X - \Delta E(p_t)X$. After the attack, the attacker keeps the value of (double spent) $X$ cryptocurrency in the standard fiat currency, $E(p_t)X$, but loses partially or fully (value of) cryptocurrency acquired for the attack, $\Delta E(p_t)X$. If $\Delta$ is sufficiently small (i.e. cryptocurrency does not collapse after the attack), then the system is vulnerable to the double-spending attack. However, if $\Delta = 1$ there is no gain from double-spending attack because the double spending attacker loses exactly as much value as (s)he gains from double spending. That is, $V_A(\Delta) = 0$ and equation (10) collapses to $[(\gamma(n_t-1)/n_t^2)(E(p_t)R_t/c_t + (\rho+\delta)q_t)]^{1/(1-\gamma)} = m_t \geq 0$. If $\Delta$ is sufficiently large, then the attack can sabotage the blockchain and lead to its complete collapse if $\Delta = 1$. In this case, the motivation of the attacker may be other than the gain (profit) from double spending (e.g. adversary power interested to damage the cryptocurrency which could include a competing centralized intermediary, a competing cryptocurrency, or other entity) (Budish 2018).

In line with Abadi and Brunnermeier (2018); Budish (2018), equation (10) implies that the equilibrium block reward to miners must be sufficiently large relative to the one-off gain from the attack. Given that the gain from the attack, $V_A(\Delta)$, is unknown (e.g. in the case of the double spending attack, $X$ an thus $V_A(\Delta) = E(p_t)X - \Delta E(p_t)X$ could be large for $\Delta < 1$) and its value might be substantial, the equilibrium mining intensity needs to be larger than the one implied by equation (7) in order to deter an attack. This is induced by the fact that the payoff from the blockchain attack, $V_A$, does not affect the economic behavior (incentives) of honest miners in allocating their computing capacity for mining (i.e. $V_A$ does not enter in equation (7)).

### 3.3 Testable hypotheses

From equations (7)-(10), we can derive three empirically testable hypotheses:

- *Mining reward hypothesis*: Security outcomes of the PoW-blockchain and the cryptocurrency price. When agents anticipate the value of cryptocurrency to be low, miners have little incentive to invest in computational resources, and the security of the network is low. The opposite is true when agents anticipate the value of cryptocurrency to be high. *Ceteris paribus, the blockchain security is sensitive (elastic) to the mining reward.*
- *Mining cost hypothesis*: The physical resource cost to write on the PoW-blockchain is intrinsically linked to the cost of preventing attacks; the security of blockchain is structurally linked to the ledger's security budget and mining costs. *Ceteris paribus, the blockchain security is sensitive (elastic) to mining costs.*

- *Mean-reverting hypothesis*: The mean-reverting behavior of the PoW-blockchain security implies that temporary cryptocurrency price shocks and mining cost shocks do not affect the long-run blockchain security. *Ceteris paribus, the PoW-blockchain security reverts back to mean in the long-run.*

**4 Estimation strategy**

**4.1 Empirical PoW blockchain security model**

The theoretical analysis established interdependencies between blockchain security, cryptocurrency market outcomes and resources devoted to the blockchain mining. Equation (7) implies that the security (measured by the allocated computing capacity) of the PoW-blockchain depends on mining rewards, the intensity of miners' competition, mining costs, discount rate and the computer equipment cost-efficiency.

In this section, we assess empirically the interdependencies between mining costs, mining rewards and the PoW-blockchain security outcomes on cryptocurrency market outcomes and mining resources. For the sake of tractability, it is useful to apply a logarithmic transformation to equation (7), which yields the following equilibrium relationship:

(11) $$y_t = b_0 + \beta x_t + u_t$$

where *y* represents the dependent variable – the PoW blockchain security (computing capacity devoted to mining), $\beta$ is a vector of coefficients to be estimated, *x* is a vector of explanatory covariates – mining rewards, $p_t R_t$, the number of miners, $n_t$, the intensity of miners' competition, $(n_t - 1)/n_t^2$, the cost of mining (including the discount rate), $c_t + (\rho + \delta)q_t$ and the computer equipment efficiency, $\varepsilon_t$, and $u_t$ is an error term.

The expected signs of coefficients for the mining reward and the intensity of miners' competition in equation (11) are expected to be positive (*number of miners* and *mining reward effects* in Figure 2). The sign of coefficient associated with the cost of mining (energy costs and discount rate) is expected to be negative (*mining cost effect* in Figure 3). The computer equipment cost-efficiency coefficient is expected to have a positive relationship with the blockchain computing capacity, because everything else constant, higher computing efficiency implies that less energy is needed to achieve a certain computing hash rate. Our primary interest is on coefficients associated with the mining reward and the cost of mining: the first coefficient measures the elasticity of the PoW-blockchain security (mining network hash rate) with respect to the mining reward (Mining reward hypothesis) and the second one with the mining costs (Mining cost hypothesis). They reflect the level of endogeneity of the security of the PoW-blockchain with respect to cryptocurrency market outcomes.

**4.2 Estimation issues**

The estimation of interdependencies between mining costs, mining rewards and blockchain security determined in equation (11) is subject to several econometric issues. The first aspect to consider is the problem of endogeneity. The endogeneity issue is particularly relevant for distributed digital ledger series, as the security outcomes of the PoW-blockchain can be determined concurrently with the cryptocurrency mining reward. For example, when distributed agents anticipate the value of cryptocurrency to be low, miners are not motivated to invest in computational resources, and the security of the blockchain would be low. In that case, crypto-coin users do not wish to accumulate large real balances, and the resulting market valuation for cryptocurrency would be low. The opposite would be true if the value of cryptocurrency is expected to be high.

To address the endogeneity problem, we rely on the Autoregressive Distributed Lag (ARDL) methodology that is being increasingly used for studying cryptocurrencies (e.g. Bouoiyour and Selmi 2015) and financial markets more generally (e.g. Stoian and Iorgulescu 2020). The ARDL bounds testing approach developed by Pesaran and Shin (1999) is particularly appropriate for estimation of the blockchain security equilibrium relationship (11) as it enables to model the long- and short-run relationships simultaneously and has several advantages over the standard cointegration methods. A key advantage for our analysis is that the ARDL approach allows treating all the relevant moments of blockchain series as potentially endogenous. As noted by Pesaran and Shin (1999, p. 16), the use of ARDL is well suitable to address the endogeneity problem: ''*appropriate modification of the orders of the ARDL model is sufficient to simultaneously correct for residual serial correlation and the problem of endogenous regressors*''.

In the context of cryptocurrencies, another important advantage is that the ARDL approach permits different number of lags for each series. Contrary to other cointegration techniques (see Engle and Granger, 1987; Phillips and Ouliaris, 1990; Johansen, 1991), the ARDL methodology does not require testing for the order of integration; it can be applied irrespective of whether the regressors are purely I(0), purely I(1) or mutually cointegrated variables (Pesaran et al., 2001). However, as pointed out by Ouattara (2004), if I(2) variables are present in the data, the computed *F* statistics of Pesaran et al. (2001) become invalid. To make sure that none of the variables is integrated of order I(2) or beyond, we investigate the integration status of the series by using the augmented Dickey–Fuller (ADF) test, the Dickey–Fuller GLS test (DF-GLS) and Phillips–Perron (PP) test. In order to find the appropriate number of lags for the series we follow the Akaike Information Criterion. Accordingly, the role of the cryptocurrency mining reward and the proof-of-work cost for each of the respective moments can be estimated after accounting for the information embedded in the lags of the entire distribution of blockchain security outcomes.

Second, there is also a potential errors-in-variables problem because part of the series is obtained from primary non-harmonized data sources and it is not straightforward to judge the reliability of these series. This concerns mainly the series that are not recorded on blockchain, (e.g. mining cost data). Indeed, the mining unit costs time series for different world regions are collected by using different sampling methodologies and different weights. These issues can be partially addressed by first differencing the data. Nevertheless, part of potential errors-in-variables issues remain. To address the remaining potential errors-in-variables, we create alternative proxies for the dependent variable – blockchain security – and key explanatory variables – mining competition – and estimate these otherwise identical mining models for robustness. The robustness checks results do not indicate any abnormal deviations in the estimated coefficients when cointegrating alternative proxies for the critical series.

**4.3 Econometric strategy**

The ARDL procedure involves two steps. First, we check for the existence of a long-run relationship by comparing the calculated F-statistic with the critical value tabulated by Pesaran et al. (2001). We begin with the general form of an ARDL(*p, q*) model:

(12) $$y_t = b_0 + \sum_{i=1}^{p} \phi y_{t-1} + \sum_{i=0}^{q} \beta_i x_{t-1} + u_t$$

where *y* represents the dependent variable – security (computing capacity) of mining, *x* is a vector of independent variables – mining rewards, intensity of miners' competition, energy costs, discount rate and the computer equipment efficiency, $b_0$ is the intercept, *p* is the number of optimal lags of the dependent variable and *q* represent the number of optimal lags of each explanatory variable.

Pesaran et al. (2001) proposed two types of critical values for a given significance level. The first type assumes that all variables in the model are I(1), whereas the second one assumes that all series are I(0). If the computed *F* statistic is below the lower bound, the null hypothesis of no long-run relationship fails to be rejected. In such case, an ARDL model in first differences without an error correction term should be estimated. If the *F*-statistic lies between the two bounds, the result is inconclusive. And finally, if the computed *F*-statistic exceeds the upper bound, the null hypothesis of no cointegration is rejected. In this case, the error correction model to be estimated is:

(13) $$\Delta y_t = b_0 + \alpha(y_{t-1} - \theta x_t) + \sum_{i=1}^{p-1} \psi_{yi} \Delta y_{t-i} + \sum_{i=0}^{q-1} \psi_{xi} \Delta x_{t-i} + u_t$$

where *θ* represent the long-run coefficients, Δ is the first difference operator, *ψ* are short-run multipliers and *α* shows the speed of adjustment of the dependent variable to a short-term shock. It measures how quickly the blockchain security adjusts to deviations from the equilibrium (Mean-reverting hypothesis).

**4.4 Specification tests**

Following the standard approach in the literature (Pesaran et al. 2001), we apply a set of diagnostic tests, as the validity of ARDL results is based on the assumption of normally distributed error terms, no serial correlation, heteroscedasticity and stability of the coefficients. The empirically estimable model specifications and the number of lags is determined based on the results from diagnostic tests, i.e. Breusch-Godfrey LM test and Durbin's alternative test for autocorrelation, Breusch-Pagan/Cook-Weisberg test for heteroscedasticity, normality testing and cumulative sum test for the parameter stability.

**5 Data**

In empirical estimations, we use Bitcoin daily data for the period 27/12/2014 – 10/01/2021. The details of data series used in estimations and their sources are reported in Table 1. All time-series are transformed in a log-form in the estimations, implying that the estimated coefficients can be interpreted as elasticities. Table 2 provides a descriptive statistic of the data used. The construction of dependent and explanatory variables is explained in the following.

Our principal data source is blockchair.com that contains records for the entire Bitcoin mining history starting from 2009 until latest transactions in 2021. For each block successfully mined, blockchair.com contains 36 block-specific characteristics: *block_id, hash, time, median_time, size, stripped_size, weight, version, version_hex, version_bits, merkle_root, nonce, bits, difficulty, chainwork, coinbase_data_hex, transaction_count, witness_count, input_count, output_count, input_total, input_total_usd, output_total, output_total_usd, fee_total, fee_total_usd, fee_per_kb, fee_per_kb_usd, fee_per_kwu, fee_per_kwu_usd, cdd_total, generation, generation_usd, reward, reward_usd, miner.* We aggregate single blocks into daily mining output, to align with the rest of the data.

**5.1 Dependent variable**

The dependent variable is the PoW-blockchain security. According to the theoretical model, the ledger security reflects the probability of an attack; a high security implies that it should be difficult for an attacker to manipulate historical or/and new records. The probability of an attack is determined by the balance of the computing power between potential attackers and honest miners. More (honest) miners and higher computing capacity imply smaller probability of a successful attack (Figure 2).

In the empirical analysis, we measure the blockchain computing capacity devoted to mining by *hash rate*, it is expressed in average daily hashes per second. According to CoinMetrics, there are several drawbacks with the hash rate index.[13] The most important one relates to the random block generation process, because of which the implied hash rate tends to follow an oscillating pattern. On the one hand, there is randomness as to whether or not a contract would settle at the top or bottom of an oscillation, which could significantly impact the outcome of a transaction. On the other hand, the hash rate can be manipulatable by large miners that control significant portions of the network hash rate. To circumvent these issues, we use *difficulty* as an alternative proxy for measuring the blockchain computing capacity. The alternative dependent variable mining *difficulty* measures the effort required to mine a new block on the blockchain. Both proxies for the network security – *hash rate* and *difficulty* – have been extracted from bitinfocharts.com (see Table 1). Both series were verified against data from blockchair.com.

---
[13] https://coinmetrics.io/coin-metrics-state-of-the-network-issue-49

### 5.2 Explanatory covariates

*Mining reward*. PoW-blockchain mining incentives are ensured via rewards for a correct and secure record keeping. The reward for every block is allocated to the miner that first solves the computational problem (hash function), by using guess and check algorithms based on the new and previous blocks of transactions. The mining reward of distributed ledgers is endogenous and fluctuates over time (in a fiat currency nomination – see Figure 1), implying that the underlying institutional governance technology may be contingent on the mining reward. In the empirical analysis, the variable mining reward is measured as the average daily value of the reward per block calculated by dividing the total mining reward per day (in US dollars) by the total number of blocks per day. Both variables – the total mining reward/day and the total number of blocks/day – have been extracted from blockchair.com (see Table 1).

*Proof-of-work costs*. Free entry and competition ensure that distributed ledgers can be more efficient and transactions less costly than centralized ledgers. As determined in the theoretical model (section 3), blockchain miners can enter freely, meaning that any agent who wishes to write on the ledger may do so by following an agreed set of rules. However, free entry of anonymous record-keepers is 'trustless' and thus requires a trust-enhancing mechanism. PoW-blockchains solves the trust problem by forcing record-keepers to pay a physical resource cost to record information and requiring that future record-keepers validate those reports. The physical resource cost to write on the blockchain is the main the cost of operating a distributed digital ledger and forms the PoW-blockchain's security budget.

In the empirical analysis, we construct a separate resource cost proxy for each global world region to measure the variable mining unit costs. According to Ciaian et al. (2021b), electricity costs account for 94-97 percent of variable mining costs of PoW blockchains. Hence, we use electricity prices in Europe (*electricity Europe*), China (*electricity China*) and North America (*electricity N. America*) to measure a region-specific cost of mining. These series have been constructed from three distinct sources: European Electricity Index (epexspot.com), Chengdu's Usage Price Electricity for Industry (ceicdata.com) and Electricity Price in North America (reports.ieso.ca) (see Table 1). To address potential errors-in-variables issues, we construct alternative proxies for measuring the variable mining unit costs and estimate these otherwise identical mining models for robustness.

As regards the fixed costs of mining, the mining *equipment efficiency* is proxied with the most efficient mining hardware available in each time period measured by the energy efficiency of the hardware (see Table 6 for an overview). This approach follows closely the literature (Zade and Myklebost 2018; CBEI 2021). We proxy the discount rate with the US 10-year treasury constant maturity rate (*10-year-treasury*). The 10-Year Treasury Constant Maturity Rate (DGS10) is extracted from fred.stlouisfed.org (see Table 1).

*Number of miners*. The total number of blockchain miners affects the blockchain security both directly and indirectly via network externalities (see Figures 2 and 3). When miners engage in the mining of blockchains, two types of opposite network externalities of the blockchain security arise, one positive and one negative. The positive network externality implies that the blockchain security is increasing with the number of miners, because each additional node reinforces the chain's security, by making it harder for any individual miner to launch an attack or to guess who will be the winning miner (Waelbroeck 2018). The negative network externality occurs because each individual miner invests in the mining-computing power, which increases both the individual miner's marginal income though also mining costs, as the difficulty of the computational problem increases in the number of miners and their computing power ("hash-power"). Increasing the difficulty of mining reduces the incentives for mining and – in the presence of learning by mining – increases the concentration of mining activities, reducing in such a way the blockchain security (Parra-Moyano, Reich and Schmedders 2019). When many small miners enter the blockchain network, likely, the positive network externality will dominate and the blockchain security outcomes will be superior compared to a highly skewed distribution of computing power across miners (few mining pools having a large share of the total network hash rate).

In the empirical analysis, we compute the total *number of miners* from the blockchair.com (see Table 1). Note that using Blockchair data we are able to distinguish between the total number of active miners and successful miners in every period. Although, the two series are correlated, their moments are different – the speed of adjustment to exogenous input and output price shocks is different between the two series. Distinguishing between the total number of active miners and successful miners is an important innovation compared to previous studies, Parra-Moyano, Reich and Schmedders (2019) is the only study we are aware of that uses a comparable decomposition technique.

C*ompetition intensity*. As discussed in section 2 and derived in the theoretical model, PoW-blockchain-based distributed record-keeping systems allow for competition: there is a free entry (every miner can write on the ledger, subject to network rules) and switching between ledgers (e.g. 'forks') is costless for users.[14] There is also competition between a potential attacker(s) and all honest miners. Incentives to record honestly make it costly for a potential attacker to distort the ledger. In the empirical analysis, we consider two alternative proxies for the competition intensity – Herfindahl-Hirschman index (*hhi*) and normalized Herfindahl-Hirschman index (*hhi normalised*) – in order to account for the unequal distribution of the blockchain computing capacity between different miners. Both Herfindahl-Hirschman concentration indices are computed based on the number of miners and the network hashrate. Both series – the total number of active miners and the network hashrate – have been extracted from blockchair.com (see Table 1).

## 6 Results

Before proceeding with the ARDL bounds testing we determine the order of integration of the variables. The test results summarized in Table 4 indicate that there are no variables integrated of the second order, which validates the use of the ARDL approach.

Table 3 summarizes the three estimated mining models with alternative specification of explanatory variables and for each of the 3 models we include 2 sub-models with alternative measures of the PoW-blockchain security, i.e. *hash rate* and *difficulty*. The three estimated mining models differ by the proxy measuring the computer intensity. Model 1 uses *competition intensity* variable, $(n_t - 1)/n_t^2$, as derived in equation (7), whereas models 2 and 3 use the two alternative proxies for competition intensity: the Herfindahl-Hirschman index (*hhi*) and the normalized Herfindahl-Hirschman index (*hhi normalised*), respectively. The rest of variables are uniform across all estimated mining models.

### 6.1 Mining reward and PoW-blockchain security

The mining reward hypothesis says that, ceteris paribus, the blockchain security is sensitive (elastic) to the mining reward. The long-run ARDL estimates tend to confirm a structural relationship between the mining reward and security outcomes of the PoW-blockchain (Table 4). This holds for both security variables measuring the blockchain computing capacity, *hash rate* and *difficulty*, and across all estimated models. The estimated elasticities of the *mining reward* variable range from 1.38 to 1.85, indicating an elastic response in the blockchain computing capacity to permanent changes in the mining reward: 1% permanent increase in the mining reward increases the underlying blockchain security by 1.38% to 1.85% in the long-run. Hence, our estimates fail to reject the mining reward hypothesis: the PoW-blockchain security is overly sensitive (elastic) to the cryptocurrency mining reward. A change in the payoff from mining causes more than proportionate change in the PoW-blockchain security.

---

[14] For example, a hard fork preserves all of the data in the parent blockchain: e.g. Bitcoin Gold and Bitcoin Cash in the case of hard forks of Bitcoin.

As discussed in section 2 and derived in the theoretical model, given that mining costs are incurred in standard fiat currencies in most cases (e.g. US dollar, Euro), the value of the mining reward fluctuates with the price of cryptocurrency,[15] which in turn affects the mining reward and mining incentives. Thus, if the expected cryptocurrency price decreases, lower mining incentives reduce the equilibrium computer mining capacity and hence the cryptocurrency security. Our estimates also imply that the reverse is valid in the case of a cryptocurrency price increase.

As regards the short-run estimates for the mining reward, they are less significant across the estimated ARDL models than the long-run results and the estimated elasticity is rather small (Table 5). A 1% positive shock in the Bitcoin mining reward (the third lag) decreases the blockchain computing capacity in the short-run by between 0.01% and 0.02%. Generally, also the short-run estimates tend to support the mining reward hypothesis: although an inverse relationship is found in our data, the security outcomes of the PoW-blockchain shows sensitivity to the Bitcoin mining reward even in the short-run. This negative relationship between the mining reward and mining intensity could be a result of other short-run effects such as mining optimization across cryptocurrencies, i.e. switching mining to other cryptocurrencies (e.g. to Bitcoin cash) when the relative price of Bitcoin to cryptocurrencies decreases.[16] This short-run inverse relationship could also be caused by secondary spiral effects induced by Bitcoin price changes – decrease (increase) – as suggested by Kroll, Davey and Felten (2013), through the subsequent loss (gain) of confidence (trust) in Bitcoin when Bitcoin mining intensity decreases (increases) which might further reduce (increase) the Bitcoin price.

**6.2 Proof-of-work cost and blockchain security**

The mining cost hypothesis says that, ceteris paribus, the blockchain security is sensitive (elastic) to mining costs. To capture a region-specific cost of mining, we have constructed distinct electricity price variables for Europe, China and the North America.

The long-run estimates for the proof-of-work cost are less significant across the estimated ARDL models than mining reward results and the estimated elasticity shows a substantial variation across world mining regions (Table 4). In line with the mining cost hypothesis, the estimated impact of variable mining unit costs is negative and statistically significant for North

---

[15] Note that the change in Bitcoin price is the main factor deriving the change in the value of mining reward because according to the algorithm the quantity of mining reward in Bitcoins, $R_t$, changes (halves) only approximately every 4 years, whereas Bitcoin price changes daily.

[16] There is some evidence of asymmetric change in Bitcoin and altcoin prices: shocks to altcoins prices tend to be greater than Bitcoin price shocks (Reiff 2018; Cheikh, Zaied and Chevallier 2020). This implies that the relative prices of Bitcoin to altcoins are inversely related with the Bitcoin price changes which may incentivize miners to shift some Bitcoin computer capacity to mining altcoins when Bitcoin price increase, and shift back the computer capacity to Bitcoin mining when Bitcoin price declines. Note that the shift in mining between different cryptos is less relevant for ASIC mining hardware, commonly used for Bitcoin mining, which is more efficient in mining specific cryptocurrencies (specific cryptographic hash algorithm) and cannot be used for mining other types of cryptocurrencies.

[16] https://www.vox.com/2019/6/18/18642645/bitcoin-energy-price-renewable-china

America and Europe. In contrast, the long-run estimates for the global mining leader China – 71.70% of the global Bitcoin hash rate are concentrated in China (Song and Aste 2020) – suggest a statistically significant and positive relationship between the proof-of-work cost and the security of the PoW-blockchain. This result is contrary to the theoretical predictions and requires some explanation.

One explanation for these geographically differentiated results could be that the intensive margin of mining is larger in China, where all major mining pools are concentrated. Given that variable mining costs are lower in China than in Europe and North America, positive shocks to electricity prices may actually increase the global share of Chinese miners. Indeed, our estimates capture other long-term behavioral effects of miners induced by a permanent change in electricity prices such as shifting mining location to places with cheaper energy (e.g. to remote regions of China, from mainland Europe to Iceland to harvest geothermal power).[17] Such long-term behavioral effects may actually increase the blockchain computing capacity, if the energy cost savings more than offset the price increase. Further, these results may also reflect the fact that the mining input cost data (which are location-specific) are less reliable than the mining reward data, which are publicly available for every single historical cryptocurrency transaction.

The short-run results for the variable mining unit costs and security outcomes of the PoW-blockchain are available for China, they cannot be examined for North America and Europe due to the estimated ARDL specifications (Table 5). In line with the theoretical model in equation (7), positive shocks to electricity prices in China have a statistically significant and negative impact on the blockchain computing capacity of the PoW-blockchain in the short-run. This result contrasts long-run estimates, where a permeant increase in electricity prices in China led to an increase in the mining intensity suggesting that other structural changes in miners' behavior might take place when the cost changes are permanent. Thus, our short-run PoW cost estimates tend support the mining cost hypothesis that the security outcomes of the PoW-blockchain is sensitive to mining costs. In the long-run, however, structural shifts and relocation of mining farms – to reduce mining operating costs – may take place and offset the short-run mining cost effect.

Overall, these ARDL bounds testing results suggest that the blockchain security is sensitive to proof-of-work costs. However, we cannot provide a definite and robust answer to the mining cost hypothesis. Instead, these results call for further analysis using more disaggregated location-specific mining cost data. Indeed, looking into proof-of-work costs and blockchain security outcomes using geographically disaggregated data offers a promising avenue for the future research.

---

[17] https://www.vox.com/2019/6/18/18642645/bitcoin-energy-price-renewable-china

A further variable capture mining costs in our model is the *hardware efficiency*. In line with the theoretical model, the hardware efficiency variable has a statistically significant positive impact on the blockchain computing capacity in all estimated models. The ARDL results imply that an increase in the efficiency of mining equipment (decrease in the input units of the computing capacity per security output unit) leads to an upgrade of security outcomes of the PoW-blockchain in the long-run. The estimated elasticities vary between 0.23 and 0.83, implying that a 1% permanent increase in the efficiency of mining equipment increases the blockchain computing capacity in the long-run by between 0.23% and 0.83%. Hence, the PoW-blockchain mining security is dependent of the mining technology available in each given point of time.[18]

### 6.3 Competition and network externalities

The theoretical mining model in (7) implies that the total blockchain computing capacity increases at a decreasing rate in the level (intensity) of competition. Our long-run estimates suggest that the miners' competition intensity (*number of miners*, *competition intensity*) has a negative impact on security outcomes of the PoW-blockchain; all long-run estimates are significantly different from zero in Table 4. These results suggest that a permanent increase in the competition intensity exercises a downward pressure on the blockchain computing capacity in the long-run.

As discussed in Section 2, digital distributed ledgers such as blockchain are subject to a number of network externalities. When new miners enter the blockchain mining, two types of direct network externalities related to the blockchain security arise, one positive and one negative. The positive network externality implies that the blockchain security is increasing with the number of miners, because each additional node reinforces the chain's security. In line with the previous literature (Waelbroeck, 2018), the negative network externality occurs because each individual miner invests in the mining-computing power, which increases both the individual miner's marginal income though also mining costs, as the difficulty of the computational problem increases in the number of miners and their computing capacity ("hash-power"). Increasing the difficulty of mining reduces the incentives for mining and – in the presence of learning by mining – increases the concentration of mining activities, which in turn reduces the blockchain security. Our estimates suggest that the negative network externality dominates of the positive network externality. Our results are in line with those of Parra-Moyano, Reich and Schmedders (2019) who find that the probability of winning a mining contest increases with the miner size. This motivates miners to join mining pools to increase their probability to win the mining contest and receive reward.[19] Indeed, our competition proxy variables are constructed based on the observed number of miners but not on the number of members within mining pools. And since a greater competition may imply fewer miners (because many individual miners join mining pools), the implied actual long-run relationship between the competition intensity and mining intensity may become negative.

---

[18] The short-run effects of electricity prices, *hardware efficiency*, *10-year-treasury* and alternative proxies for *competition intensity* cannot be examined due to the ARDL specifications, as no lags of these dependent variables entered the model.

In the short-run, the mining competition intensity has a statistically positive impact on the blockchain computing capacity in all estimated models. These results also indicate that in the short-run, the competition among miners encourages deployment of more mining capacity in line with the model derived in equation (7). While in the short-run the miners' competition leads to expansion of the blockchain mining capacity, in the long-run the inverse relationship is valid indirectly suggesting reduced competition level as individual miners have the incentive to join mining pools.

### 6.4 Dynamics and the mean-reverting of the blockchain security

The mean-reverting hypothesis says that, ceteris paribus, the PoW-blockchain security reverts back to mean in the long-run. The estimates of the error correction term – which measure the speed of adjustment of the short-run dynamics of mining to the long-run equilibrium path – are statistically significant across all models. The error correction terms vary between -0.002 and -0.009, implying that between 0.20% and -0.90% of the long-run disequilibrium in mining intensity is corrected by the short-run adjustment on the same day. In other words, the disequilibrium corrects at an average speed of convergence of between 0.20% and 0.90% per day. In terms of the duration, any deviation from the long-run equilibrium is corrected in around 109 to 447 days. These results provide support for the mean-reverting hypothesis saying that in response to shocks and short-run deviations security outcomes of the PoW-blockchain revert back to the equilibrium security level in the long-run.

The lagged dependent variable (proxied by *hash rate* and *difficulty*) is statistically significant in all estimated mining models. The coefficient estimates vary between -0.03 and -0.47. The relatively high values of these coefficients indicate that the temporary shocks in the mining computing capacity disappear over time relatively fast: in around 2 to 29 days. These results support the mean-reverting hypothesis that the security outcomes of the PoW-blockchain are sensitive to Bitcoin market outcomes in the short-run to fluctuations with instant shocks disappearing over a short time period (within few days).

The *10-year-treasury* variable, which is a proxy for the discount rate, has a statistically significant positive effect on the blockchain computing capacity. This result suggests that that the *10-year-treasury* actually captures a miner investment competition effect, i.e. miners perceive it as an alternative investment asset. As far as cryptocurrency is perceived as an investment asset, shocks to competing investment asset returns (including *10-year-treasury*) are expected to impact positively miners' choices to invest in the mining of cryptocurrency. Our results confirm that miners perceive cryptocurrency to be competing for investment with other financial assets and thus need to generate a competitive return. The return arbitrage among alternative potential investment opportunities implies a positive price relationship between cryptocurrency and alternative financial assets (Murphy 2011; Ciaian et al. 2018, Ciaian et al. 2021a). Thus, the positive coefficient associated with the *10-year-treasury* variable implies that miners are motivated to invest in more computing capacity for mining when the returns to financial assets increase.

---

[19] Other benefit of joining mining pools is that it creates a steady stream of income, rather than greater income but at lower frequency (i.e. due to lower odd of winning the mining contest) with individual mining (Liu and Wang 2017).

## 7 Discussion and concluding remarks

The present paper has studied the interdependencies between mining costs, mining rewards and blockchain security. We have attempted to answer the following questions. To what extent the cost of operating blockchains is intrinsically linked to the cost of preventing attacks? To what extent the digital ledger's record-keeping security budgets (measured by mining rewards in a fiat currency nomination) of cryptocurrencies are correlated with the cryptocurrency market outcomes? In this paper, we have focused on the proof-of-work (PoW) blockchain, which is a particularly interesting blockchain to study as the involved physical resource expenditures provide a distinct advantage in achieving consensus among distributed miners.

First, we have theoretically derived an equilibrium relationship between cryptocurrency price, mining rewards and mining costs, and blockchain security outcomes. Second, using daily Bitcoin data for 2014–2021 and employing the autoregressive distributed lag approach – that allows treating all the relevant moments of the blockchain series as potentially endogenous – we have provided empirical evidence about interdependencies between mining costs, mining rewards and blockchain security. Our results suggest that the cryptocurrency price and mining rewards are intrinsically linked to blockchain security outcomes. In contrast, the physical resource cost to write on the blockchain – the cost of operating the PoW-blockchain – is only weakly cointegrated with the strength of the network security; the ARDL results for mining costs are geographically differentiated, implying heterogeneities in variable mining costs across global world mining regions.

Our main contribution to the literature is formally establishing a link between the probability distribution over security outcomes that permanently depend on the underlying distribution of cryptocurrency market outcomes and providing a supporting empirical evidence. Our results complement findings of this emergent literature by quantifying how the probability distribution over security outcomes permanently depends on the underlying distribution of cryptocurrency market outcomes. Due to the extremely high cryptocurrency return volatility, the PoW-based blockchain security budget is exposed to high volatility and may result in a series of low-security equilibriums and high-security equilibriums.

**Figure 1. Bitcoin price and aggregate security spend, USD**

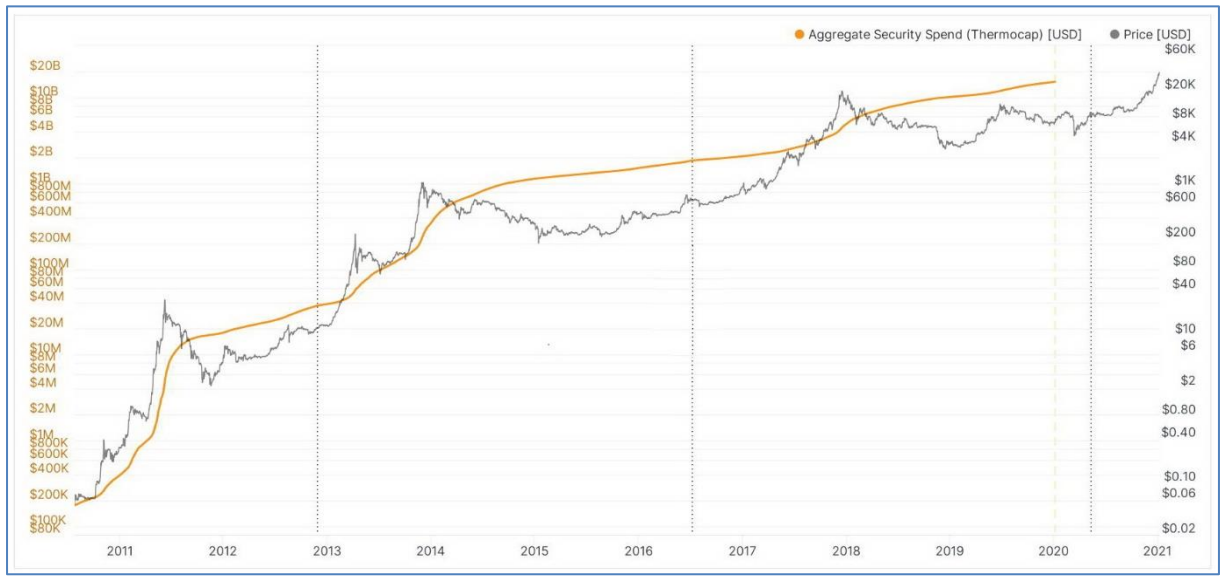

Source: coinmetric.com and studio-glass-node. Notes: Bitcoin's cumulative miner revenue – Thermocap – is calculated by taking the running sum of daily miner revenue in USD. Market capitalization to Thermocap provides an indication of the Bitcoin's current market value compared to the aggregate amount spent to secure the network. Note that, due to lags in the adjustment of mining investment, Thermocap is relatively slow moving and does not have the same level of volatility as the market capitalization.

**Figure 2. Interdependencies between bitcoin price and blockchain security**

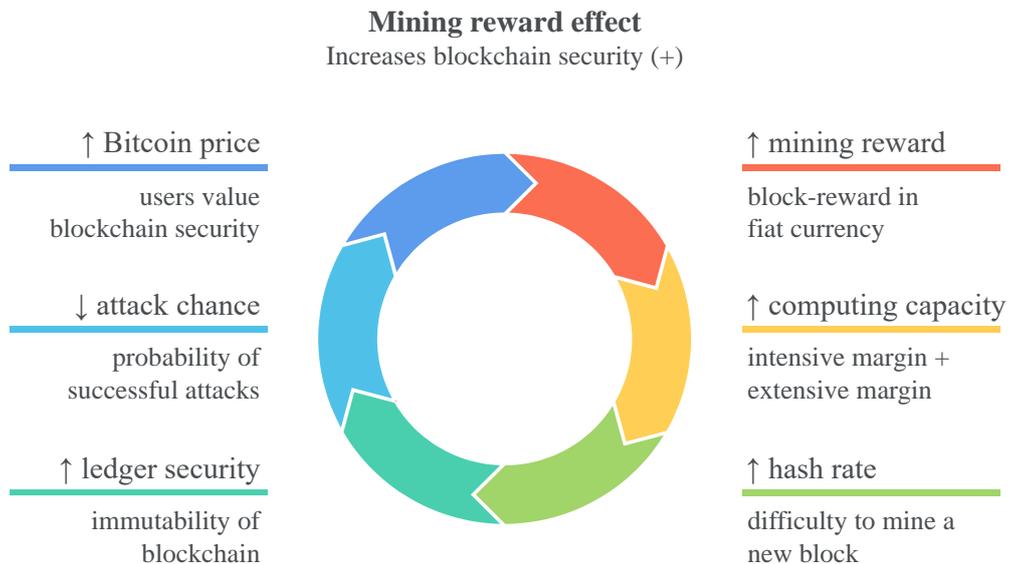

Source: Conceptual framework (section 3).

# Figure 3. Interdependencies between proof-of-work cost and blockchain security

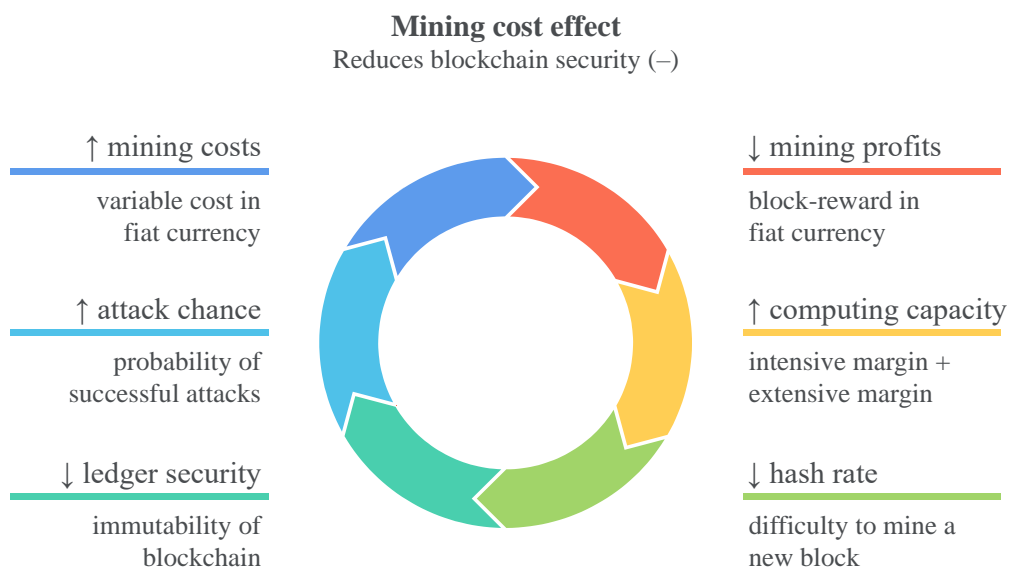

**Mining cost effect**
Reduces blockchain security (–)

↑ mining costs — variable cost in fiat currency
↓ mining profits — block-reward in fiat currency
↑ attack chance — probability of successful attacks
↑ computing capacity — intensive margin + extensive margin
↓ ledger security — immutability of blockchain
↓ hash rate — difficulty to mine a new block

Source: Conceptual framework (section 3).

# Table 1. Data sources

| Variable | Unit | Description of variable | Source |
| --- | --- | --- | --- |
| **Dependent variable** | | | |
| hash rate | Hash/second | Total computing capacity | bitinfocharts.com |
| difficulty | Average difficulty per day | Mining difficulty | bitinfocharts.com |
| **Explanatory variables** | | | |
| mining reward | USD per block | Mining reward per block (*reward_usd/ no_bl_total*) | blockchair.com |
| PoW cost: electricity Europe | EUR/MWh | European Electricity Index | www.epexspot.com |
| PoW cost: electricity China | USD/kWh | Chengdu's Usage Price Electricity Industry, USD | www.ceicdata.com |
| PoW cost: electricity N. America | CAD/MWh | Electricity price in North America | ieso.ca |
| hardware efficiency | J/Giga hash | Mining equipment efficiency – Bitcoin mining hardware generation (the most efficient device in each period) | Constructed based on: Zade and Myklebost (2018), CBEI (2019) |
| number of miners | No | Number of miners, $n_t$ | blockchair.com |
| competition intensity | Index | Competition intensity | computed $(n_t - 1)/n_t^2$ |
| hhi | Index | Herfindahl-Hirschman index | computed based on $n_t$ and hashrate |
| hhi normalised | Index | Normalised Herfindahl-Hirschman index | computed based on on $n_t$ and hashrate |
| 10-year-treasury | % | 10-Year Treasury Constant Maturity Rate (DGS10) | fred.stlouisfed.org |

## Table 2. Descriptive statistics of used data

| Variable | Obs | Mean | Std. Dev. | Min | Max |
|---|---|---|---|---|---|
| *Dependent variable* | | | | | |
| hashrate | 2207 | 40.612 | 6.475 | 25.442 | 52.332 |
| difficulty | 2207 | 24.028 | 6.923 | 9.581 | 32.709 |
| *Explanatory variables* | | | | | |
| mining_reward | 2207 | 9.954 | 2.387 | -0.692 | 14.713 |
| PoW cost: electricity Europe | 2207 | 3.961 | 0.981 | -6.908 | 5.429 |
| PoW cost: electricity China | 2207 | -2.506 | 0.067 | -2.402 | -2.409 |
| PoW cost: electricity N. America | 2207 | 2.809 | 2.177 | -6.908 | 6.010 |
| hardware efficiency | 2207 | 0.172 | 3.090 | -3.219 | 6.552 |
| number of miners | 2207 | 3.068 | 0.934 | 0.000 | 3.984 |
| competition intensity | 2207 | -3.301 | 1.089 | -6.908 | -1.497 |
| hhi | 2207 | -1.851 | 0.770 | -2.608 | 0.000 |
| hhi normalised | 2207 | -2.173 | 0.946 | -3.176 | 0.000 |
| 10-year-treasury | 2207 | 0.893 | 0.229 | 0.315 | 1.428 |

## Table 3. Specification of empirical models

| | Dependent variable: hashrate | | | Dependent variable: difficulty | | |
|---|---|---|---|---|---|---|
| | M1.1 | M2.1 | M3.1 | M1.2 | M2.2 | M3.2 |
| *Dependent variable* | | | | | | |
| hashrate | X | X | X | | | |
| difficulty | | | | X | X | X |
| *Explanatory variables* | | | | | | |
| mining reward | X | X | X | X | X | X |
| PoW cost: electricity Europe | X | X | X | X | X | X |
| PoW cost: electricity China | X | X | X | X | X | X |
| PoW cost: electricity N. America | X | X | X | X | X | X |
| hardware efficiency | X | X | X | X | X | X |
| number of miners | X | X | X | X | X | X |
| competition intensity | X | | | X | | |
| hhi | | X | | | X | |
| hhi normalised | | | X | | | X |
| 10-year-treasury | X | X | X | X | X | X |

## Table 4. Estimation results: long-run interdependencies

| | Dependent variable: hashrate | | | Dependent variable: difficulty | | |
|---|---|---|---|---|---|---|
| | M1.1 | M2.1 | M3.1 | M1.2 | M2.2 | M3.2 |
| mining reward | 1.398*** | 1.428*** | 1.379*** | 1.703*** | 1.848*** | 1.845*** |
| PoW cost: electricity Europe | -0.236** | -0.331** | -0.340** | -0.137* | -0.155* | -0.183* |
| PoW cost: electricity China | 1.716** | 1.206** | 2.568** | 1.238* | 2.857* | 3.108* |
| PoW cost: electricity N. America | -0.132** | -0.214** | -0.240** | -0.061* | -0.120* | -0.142* |
| hardware efficiency | 0.839*** | 0.546*** | 0.650*** | 0.494*** | 0.237*** | 0.372*** |
| number of miners | -1.154*** | -2.824*** | -2.487*** | -2.112*** | -4.027*** | -3.393*** |
| competition intensity | -1.143*** | | | -1.257*** | | |
| hhi | | -3.596** | | | -5.756*** | |
| hhi normalised | | | -2.285*** | | | -4.595*** |
| 10-year-treasury | 2.204** | 4.869** | 5.994* | 2.556* | 5.675** | 7.364* |
| *Error correction term* | | | | | | |
| hash rate (-1) | -0.009*** | -0.008*** | -0.007*** | | | |
| difficulty (-1) | | | | -0.003** | -0.004** | -0.002** |
| Speed-of-adjustment (days) | 109 | 121 | 135 | 317 | 350 | 447 |

Notes: Speed-of-adjustment is calculated based on the error correction rate. ***significant at 1% level, **significant at 5% level, *significant at 10% level. Empty cells indicate absence of a variable in the respective model.

**Table 5. Estimation results: short-run interdependencies**

|  | Dependent variable: hashrate | | | Dependent variable: difficulty | | |
|---|---|---|---|---|---|---|
|  | M1.1 | M2.1 | M3.1 | M1.2 | M2.2 | M3.2 |
| Δ dependent variable (-1) | -0.444*** | -0.441*** | -0.446*** | 0.192*** | 0.214*** | 0.202*** |
| Δ dependent variable (-2) | -0.469*** | -0.452*** | -0.431*** | -0.146*** | -0.148*** | -0.148*** |
| Δ dependent variable (-3) | -0.317*** | -0.293*** | -0.291*** | -0.039*** | -0.034*** | -0.034*** |
| Δ dependent variable (-4) | -0.230*** | -0.196*** | -0.203*** | -0.075*** | -0.071*** | -0.073*** |
| Δ dependent variable (-5) | -0.164*** | -0.143*** | -0.152*** | -0.058** | -0.057*** | -0.056*** |
| Δ dependent variable (-6) | -0.099** | -0.096*** | -0.092*** | -0.053** | -0.047*** | -0.049** |
| Δ dependent variable (-7) | -0.074** | -0.064** | -0.064** | -0.061** | -0.057** | -0.051** |
| Δ mining reward | -0.019*** | -0.017*** | -0.017*** | -0.001*** | -0.001*** | -0.001*** |
| Δ mining reward (-1) | -0.020** | -0.034** | -0.016** | -0.003** | -0.002* | -0.002* |
| Δ mining reward (-2) |  |  |  | 0.001* | 0.001* | 0.001* |
| Δ mining reward (-3) |  |  |  | -0.014* | -0.015 | -0.013* |
| Δ mining reward (-4) |  |  |  | -0.015 | -0.015 | -0.015 |
| Δ PoW cost: electricity China | -1.258*** | -1.319*** | -1.181*** |  |  |  |
| Δ number of miners |  |  |  |  | 0.002** | 0.003** |
| Δ number of miners (-1) |  |  |  |  | 0.077 | 0.076 |
| Δ number of miners (-2) |  |  |  |  | 0.092 | 0.094 |
| Δ competition intensity | 0.010** |  |  | 0.001** |  |  |
| Δ competition intensity (-1) | 0.063** |  |  | 0.033* |  |  |
| Δ competition intensity (-2) | 0.041* |  |  | 0.033 |  |  |
| constant | 0.292** | 0.350** | 0.383** | 0.046** | 0.074** | 0.103** |

Notes: ***significant at 1% level, **significant at 5% level, *significant at 10% level. Empty cells indicate either absence of a variable in the respective model or the coefficient or the variable is not selected in the estimation; Δ is difference.

**Table 6. Development of the PoW mining hardware efficiency**

| Type | Hardware name | Date | J/Th |
|---|---|---|---|
| CPU | ARM Cortex A9 | 3 Oct 2007 | 877,193 |
| GPU | ATI 5870M | 23 Sep 2009 | 264,550 |
| FPGA | X6500 FPGA Miner | 29 Aug 2011 | 43,000 |
| ASIC | Canaan AvalonMiner B1 | 1 Jan 2013 | 9,351 |
| ASIC | KnCMiner Jupiter | 5 Oct 2013 | 1,484 |
| ASIC | Antminer U1 | 1 Dec 2013 | 1,250 |
| ASIC | Bitfury BF864C55 | 3 Mar 2014 | 500 |
| ASIC | RockerBox | 22 Jul 2014 | 316 |
| ASIC | ASICMiner BE300 | 16 Sep 2014 | 187 |
| ASIC | BM1385 | 19 Aug 2015 | 181 |
| ASIC | PickAxe | 23 Sep 2015 | 140 |
| ASIC | Antminer S9-11.5 | 1 Jun 2016 | 98 |
| ASIC | Antminer R4 | 1 Feb 2017 | 97 |
| ASIC | Ebang Ebit 10 | 15 Feb 2018 | 92 |
| ASIC | 8 Nano Compact | 1 May 2018 | 51 |
| ASIC | Antminer S17 | 9 Apr 2019 | 36 |
| ASIC | Antminer S19 Pro | 23 Mar 2020 | 30 |

Source: Song and Aste (2020)